\documentclass[twocolumn,showpacs,preprintnumbers,amsmath,amssymb]{revtex4}

\usepackage{graphicx}
\usepackage{dcolumn}
\usepackage{bm}

\newcommand{\Eq}[1]{Eq.~({\protect\ref{#1}})}

\newcommand{\Ref}[1]{Ref.\protect\cite{#1}}
\newcommand{\Refs}{Refs.}
\newcommand{\Sect}[1]{Sect.~\protect\ref{#1}}
\newcommand{\Appendix}[1]{Appendix.~\protect\ref{#1}}
\newcommand{\Fig}[1]{Fig.~\protect\ref{#1}}

\newcommand{\Table}[1]{Table~\protect\ref{#1}}

\newcommand{\Tate}{\rule{0cm}{1.1em}}
\newlength{\Tatescale}
\newcommand{\Hs}{\hspace*{1em}}
\newcommand{\Bs}{\hspace*{-0.5em}}

\newcommand{\zr}[1]{\mbox{\hspace*{#1em}}}

\newcommand{\ZZ}{\mbox{\sf Z\zr{-0.45}Z}}

\newlength{\figwidth}
\newcounter{subfigure}
\newcommand{\Cut}[1]{}
\begin{document}
\title{
Penta-quark baryon in anisotropic lattice QCD
}
\author{
  N.~Ishii$^1$ \footnote{E-mail : ishii@rarfaxp.riken.jp},
  T.~Doi$^2$,
  H.~Iida$^1$,
  M.~Oka$^1$,
  F.~Okiharu$^3$,
  and H.~Suganuma$^1$
}
\affiliation{$^1$
  Department of Physics, H-27,
  Faculty of Science, Tokyo Institute of Technology,
  Meguro, Tokyo 152-8551, Japan
}
\affiliation{$^2$
  RIKEN BNL Research Center, Brookhaven National Laboratory,
  Upton, New York 11973, USA
}
\affiliation{$^3$
  Department of Physics, Faculty of Science and Technology, Nihon University,
  1-8-14 Kanda-Surugadai, Chiyoda, Tokyo 101-8308, Japan
}


\begin{abstract}
The penta-quark(5Q) baryon is  studied in anisotropic quenched lattice
QCD with renormalized anisotropy $a_s/a_t$=4 for a high-precision mass
measurement. The standard Wilson action at $\beta=5.75$ and the $O(a)$
improved  Wilson quark  action with  $\kappa$=0.1210(0.0010)0.1240 are
employed  on a  $12^3  \times 96$  lattice.   Contribution of  excited
states is suppressed by using a smeared source.
We investigate both the  positive- and negative-parity 5Q baryons with
$I=0$ and spin $J=1/2$ using a non-NK-type interpolating field.
After chiral extrapolation, the  lowest positive-parity state is found
to have a mass, $m_{\Theta}=2.25$  GeV, which is much heavier than the
experimentally observed  $\Theta^+(1540)$.  The lowest negative-parity
5Q  appears at  $m_{\Theta}=1.75$ GeV,  which  is near  the s-wave  NK
threshold.  To  distinguish spatially-localized 5Q  resonances from NK
scattering states, we  propose a new general method  imposing a ``{\it
Hybrid  Boundary  Condition  (HBC)}'',   where  the  NK  threshold  is
artificially raised without affecting compact five-quark states.
The study  using the HBC  method shows that the  negative-parity state
observed  on  the   lattice  is  not  a  compact   5Q  but  an  s-wave
NK-scattering state.
\end{abstract}

\maketitle
\section{Introduction}
Quantum  chromodynamics  (QCD) is  the  fundamental  theory of  hadron
physics.  It specifies elementary  interactions at the level of quarks
and gluons,  which serve  as building-blocks to  construct complicated
interactions among hadrons.  At high energy, the QCD coupling constant
diminishes due to asymptotic freedom, and perturbative QCD works well.
At low energy, however, with  the growth of the QCD coupling constant,
perturbation  theory  breaks down,  and  the  system  is dominated  by
non-perturbative effects.

One of  the most important  features of non-perturbative QCD  is color
confinement,   i.e.,  all   hadrons  are   formed  as   color  singlet
combinations  of  quarks and  gluons.   In  the conventional  picture,
mesons consist of a quark and  an antiquark, while baryons are made of
three quarks.  QCD, however, allows more general hadrons, often called
exotic hadrons, which  contain extra pairs of quark  and antiquark, or
gluons. Possible candidates include multiquark hadrons, glueballs, and
hybrid  hadrons.   Experimental searches  found  no manifestly  exotic
hadrons  before  the  year  2002.   Therefore, the  discovery  of  the
manifestly exotic  narrow resonance $\Theta^+(1540)$ by  LEPS group at
SPring-8   was  an   epoch  making   event  in   the   hadron  physics
\cite{nakano}.

$\Theta^+$ is confirmed to have baryon number $B=1$, charge $Q=+1$ and
strangeness $S=+1$.  It has to contain at least one anti-strange quark
and   its  simplest   configuration   is  $uudd\bar{s}$.    Therefore,
$\Theta^+$ is a manifestly exotic penta-quark state.

The experimental search \cite{nakano} of $\Theta^+$ was motivated by a
theoretical prediction given in \Ref{diakonov}.
It is noted  that $\Theta^+$ had been considered  several times in the
history \cite{jaffe-76, strottman, praszalowicz}.
%
%
The  discovered peak  in  the  $nK^+$ invariant  mass  is centered  at
$1.54\pm 0.01$ GeV with a width smaller than 25 MeV.
A number of  groups \cite{diana,clas,saphir,experiments} confirmed the
LEPS discovery,  among which \Ref{saphir} claims  that $\Theta^+$ must
be  isoscalar,  since  no  $\Theta^{++}$  is observed  in  the  $pK^+$
invariant mass spectrum.

Numerous  theoretical  studies of  penta-quark  baryons have  appeared
since its  discovery \cite{praszalowicz, cohen,  itzhaki, kim, hosaka,
jaffe, lipkin, carlson-positive, stancu, jennings, glozman, enyo, zhu,
matheus,   sugiyama,  carlson,   shinozaki,  huang,   maezawa,  shlee,
narodetskii,  sugamoto, suganuma,  okiharu, alexandrou,  bicudo, oset,
capstick}.
One  of the important  issues is  the spin  and parity  of $\Theta^+$.
Although $J^{\pi} = \frac1{2}^+$ is  suggested in the original work of
\Ref{diakonov} based on  the chiral soliton model, there  is no direct
experimental    information    available    on   spin    and    parity
\cite{nakano-hicks}.
An educated  guess for the  spin is $J=1/2$, since  the color-magnetic
interaction favors smaller spin states in general.
There  is,  however, no  consensus  on  parity  at all.   Experimental
determination  of  the  parity  of  $\Theta^+$  happens  to  be  quite
challenging   \cite{thomas},    while   opinions   are    divided   in
theory\cite{oka,zhu.review}.

Positive  parity is  supported  by various  model calculations,  i.e.,
soliton  models  \cite{diakonov,  praszalowicz, cohen,  itzhaki,  kim,
weigel}, chiral  bag model \cite{hosaka},  the Jaffe-Wilczek's diquark
model  \cite{jaffe},   the  Karliner-Lipkin's  diquark-triquark  model
\cite{lipkin},  some quark model  calculations \cite{carlson-positive,
stancu,  jennings,  glozman,   enyo},  and  other  model  calculations
\cite{bicudo, oset}.  Negative parity is supported by some other quark
model   calculations   \cite{carlson,   huang,  shinozaki,   jaffe-76,
strottman}, and QCD sum rules \cite{sugiyama}.

There  are  several  lattice  QCD studies  of  penta-quarks  available
\cite{sasaki,scikor,kentacky,chiu,ishii,rabbit},  but  they  have  not
reached   a  consensus   yet.   Except   for  \Ref{chiu},   all  these
calculations support that negative  parity states are lighter than the
positive parity ones and  that positive-parity penta-quarks may appear
at the mass above  2 GeV.  \Refs\cite{scikor,sasaki} claimed existence
of  a bound  negative-parity penta  quark, whereas  \Ref{kentacky} has
observed  no evidence  for narrow  resonances.  Naively,  the negative
result by \Ref{kentacky} may be understood as a natural consequence of
their choice of the NK-type  interpolating field, since it is expected
to couple strongly to the NK continuum state.
Therefore, under  these circumstances, it is important  to use various
interpolating  fields  and accumulate  more  data  to  give a  precise
prediction.

The aim  of this  paper is twofold.   (i) We provide  a high-precision
data  on the penta-quark  baryon $\Theta^+$  by using  the anisotropic
lattice  QCD.  (ii)  We  propose a  new  method, which  enables us  to
distinguish compact  penta-quarks from NK scattering  states and apply
it to  the negative-parity $\Theta^+$.

The      paper       is      organized      as       follows.       In
\Sect{section.interpolating.field},  we  discuss  our  choice  of  the
interpolating field for the  penta-quark $\Theta^+$, and introduce the
parity projection.
We  adopt the  standard Wilson  gauge  action at  $\beta=5.75$ on  the
$12^3\times  96$ lattice with  the renormalized  anisotropy $\xi\equiv
a_{s}/a_{t}  = 4$.  The  use of  the anisotropic  lattice is  known to
serve  as  a  powerful  tool  for high-precision measurements  of  
temporal
correlators \cite{klassen,matsufuru,nemoto,ishii-gb}.
For quark action, we adopt $O(a)$-improved Wilson (clover) action with
four  values  of  the  hopping parameter  as  $\kappa=0.1210  (0.0010)
0.1240$.
We also employ a smeared operator to enhance the low-lying spectra.

In \Sect{section.numerical.result.i}, we present our numerical results
for both positive and negative parity channels.
For both  cases, we observe a  rather stable plateau  in the effective
mass plot.
For   the  positive  parity   case,  we   obtain,  after   the  chiral
extrapolation, $m_{\Theta}=2.25$  GeV, which is much  heavier than the
observed $\Theta^+(1540)$.
For the  negative parity case, we obtain  $m_{\Theta}=1.75$ GeV, which
is rather close to the empirical value.
Our  data  thus  suggest  that the  negative-parity state is the ground 
state of  the
penta-quark spectrum.

To clarify  whether this negative  parity state is really  a localized
resonance or  not, we propose a  new general method with  a new ``{\em
hybrid boundary condition (HBC)}'' for the quark fields.
\Sect{section.hybrid.boundary.condition} is  devoted to description of
the HBC method.
In the HBC method, the  spatial boundary condition on the quark fields
are artificially taken so that N and K must have non-vanishing spatial
momenta, while it does not  affect localized penta-quark states.  As a
consequence, the HBC  raises the s-wave NK threshold  by a few hundred
MeV  without   changing  the  mass  of  a   compact  resonance  state,
$\Theta^+$.

By applying the HBC  method, we  investigate  the negative-parity
state further. Particularly, we examine difference of the spectra 
between
the two boundary conditions.
We find  that, in the effective  mass plots, the plateau is raised by
about the same  amount as the shift of  the NK-thresholds.
We therefore conclude that there is no localized resonance state below
$\sqrt{m_{\rm  N}^2  +  \vec{p}_{\rm  min}^2} +  \sqrt{m_{\rm  K}^2  +
\vec{p}_{\rm min}^2}$  with $|\vec p_{\rm  min}|=\sqrt{3}\pi/L$.  Thus
the state observed under the  standard boundary condition turns out to
be an NK scattering state.
%
%

In  \Sect{section.summary.and.discussion},  a  summary  is  given  and
possible implications of our results are presented.

\section{interpolating field and parity projection}
\label{section.interpolating.field}
We  consider a  non-NK type  interpolating field  for  the penta-quark
$\Theta^+$ as
\begin{equation}
  O_{\alpha}
  \equiv
  \epsilon_{abc}\epsilon_{ade}\epsilon_{bfg}
  \left( u^T_d C\gamma_5 d_e \right)
  \left( u^T_f C d_g \right)
  (C \bar{s}^T_c)_\alpha,
\label{sasaki-op}
\end{equation}
where $\alpha$ denotes  the Dirac index, and $a-g$  are color indices.
$C\equiv\gamma_4\gamma_2$ denotes charge conjugation matrix.
The  quantum number  of  \Eq{sasaki-op} is  spin  $J=1/2$ and  isospin
$I=0$.
It has an  advantage that it cannot be decomposed into  a product of N
and K  in the  non-relativistic limit. Hence,  we may expect  that its
coupling to NK scattering states is rather weak.
This is  the crucial  difference between our  calculation and  that of
\Ref{kentacky}, who adopted the  NK type interpolating field, which is
expected to couple to NK scattering states strongly.
The NK type interpolating field  would not be an appropriate choice in
studying the narrow penta-quark resonance in this sense.

We    present    Fierz    rearrangements    of    \Eq{sasaki-op}    in
\Appendix{appendix}.  We  can explicitly see that the  NK component in
the correlation  function is suppressed  by a factor 1/16  compared to
that from the NK-type operator.
This  non-NK type  interpolating field  \Eq{sasaki-op} was  adopted in
\Refs~\cite{sugiyama,sasaki,chiu}.

Under the spatial reflection,  the quark fields transform as $q(t,\vec
x) \to  \gamma_4 q(t,-\vec x)$.   Therefore, \Eq{sasaki-op} transforms
as
\begin{equation}
  O(t,\vec x)
  \to
  + \gamma_4 O(t,-\vec x).
\end{equation}
This means  that the intrinsic  parity of \Eq{sasaki-op}  is positive.
Although  the  intrinsic  parity  of \Eq{sasaki-op}  is  positive,  it
couples to negative parity states as well \cite{montvay}. In the Dirac
representation,  one sees  that  the upper  (lower) component  creates
positive   (negative)  parity   states.   In   addition,   in  lattice
formulation   with   limited  temporal   extension,   it  may   create
backwardly-propagating state  with opposite parity.   Since the parity
projection  of baryon  is thus  involved, we  explain our  strategy of
parity projection.

We  consider the  asymptotic behavior  of the  zero-momentum projected
correlator as
\begin{equation}
  G_{\alpha\beta}(t)
  \equiv
  \frac1{V}
  \sum_{\vec x}
  \langle O_{\alpha}(t,\vec x)\bar{O}_{\beta}(0,\vec 0)\rangle,
\end{equation}
in large $t$ and  large $N_t - t$ region. ($V$ and  $N_t$ refer to the
spatial volume and the temporal size of the lattice, respectively.)
In  this region,  the correlator  is assumed  to be  dominated  by the
lowest-lying states of  each parity, and is decomposed  into two parts
in the following manner \cite{montvay}:
\begin{eqnarray}
  G^{(\pm)}(t)
  &\equiv&
  P_{+}
  \left( C_+ e^{-m_+ t} \pm C_- e^{-m_-(N_t - t)} \right)
  \label{correlator}
  \\\nonumber
  &+&
  P_{-}
  \left( C_- e^{-m_- t} \pm C_+ e^{-m_+(N_t - t)} \right),
\end{eqnarray}
where the  choice of  ``$\pm$'' depends on  the boundary  condition of
quark  fields  in  imaginary  time, i.e.,  ``$+$''  for  anti-periodic
boundary condition and ``$-$'' for periodic boundary condition.
$m_+$  and $m_-$  refer  to  the energies  of  lowest-lying states  in
positive  and  negative  parity  channels, respectively.
$P_{\pm}\equiv (1  \pm \gamma_4)/2$  work as projection  matrices onto
the ``upper'' and the ``lower'' Dirac subspaces, respectively.

We consider the first term  in \Eq{correlator}.  In the region $0\ll t
\ll N_t/2$, $C_+  e^{-m_+ t}$ is expected to  dominate the correlator,
whereas, in the  region $N_t/2 \ll t \ll  N_t$, $C_-e^{-m_-(N_t - t)}$
is expected  to dominate the correlator.  In  the intermediate region,
i.e., $t \sim N_t/2$, the  contributions from the both terms cannot be
negligible.
The second  term in \Eq{correlator}  behaves in a similar  manner.
%
%
In this way, by confining ourselves  to the region $0\ll t \ll N_t/2$,
and by applying  the projection matrices $P^{(\pm)}$ to  $G(t)$, we can
pick up the positive and negative parity states separately.

One may point out that the  exact parity projection can be achieved by
summing up $G^{(+)}$ and $G^{(-)}$, because the backwardly-propagating
contributions  cancel.   This is  often  used  to  project parity  for
ordinary ``three-quark''  baryons at  quenched level. However,  in the
case of penta-quark, this may not work even in quenched level, because
the  five-quark system  can  decompose into  N  and K  so  that N  may
propagate  forward whereas  K  may propagate  backward  or vice  versa
\cite{rabbit}.

\section{Numerical result on $\Theta^+$ mass}
\label{section.numerical.result.i}
\subsection{Lattice parameter set}

\begin{table*}
\begin{ruledtabular}
\begin{tabular}{ccccccccccl}
$\beta$ & $\gamma_{\rm  G}$ & $a_s/a_t$ & $a_{s}^{-1}$  [GeV] & Size &
$N_{\rm conf}$ & $u_{\sigma}$ & $u_{\tau}$ & $\gamma_F$ & $\kappa_c$ &
Values of $\kappa$ \\
\hline
5.75 &  3.2552 & 4 &  1.100(6) & $12^3\times  96$ & 504 &  0.7620(2) &
0.9871(0) & 3.909 & 0.12640(5) & 0.1240, 0.1230, 0.1220, 0.1210
\end{tabular}
\end{ruledtabular}
\caption{Parameters  of the lattice  simulation.  The  spatial lattice
spacing $a_{\sigma}$ is  determined with $r_0^{-1} = 385$  MeV for the
Sommer parameter.  The mean-field values of link variables ($u_\sigma$
and $u_\tau$) are defined in the Landau gauge.  $\kappa_c$ denotes the
critical  value of  $\kappa$. }
\label{table.lattice.parameters}
\end{table*}
To generate gauge field  configurations, we use the standard plaquette
action on the anisotropic lattice of the size $12^3\times 96$ as
\begin{eqnarray}
  S_{\rm G}
  &=&
  \frac{\beta}{N_c}
  \frac1{\gamma_{\rm G}}
  \sum_{s,i<j\le3}
  \mbox{Re} \mbox{Tr}
  \left\{ 1 - P_{ij}(s)\right\}
  \\\nonumber
  &+&
  \frac{\beta}{N_c}
  \gamma_{\rm G}
  \sum_{s,i\le 3}
  \mbox{Re} \mbox{Tr}
  \left\{ 1 - P_{i4}(s)\right\},
\end{eqnarray}
where $P_{\mu\nu}(s) \in  \mbox{SU(3)}$ denotes the plaquette operator
in  the  $\mu$-$\nu$-plane.   The   lattice  parameter  and  the  bare
anisotropy parameter are  fixed as $\beta \equiv 2N_c/g^2  = 5.75$ and
$\gamma_{\rm    G}=3.2552$,   respectively,   which    reproduce   the
renormalized anisotropy as $\xi\equiv a_{s}/a_{t} = 4$ \cite{klassen}.
Adopting  the  pseudo-heat-bath  algorithm,  we pick  up  gauge  field
configurations every  500 sweeps after skipping 10,000  sweeps for the
thermalization.   We use  totally  504 gauge  field configurations  to
construct the temporal correlators.
The  lattice spacing  is  determined from  the  static quark  potential
adopting the Sommer parameter $r_0^{-1} = 395$ MeV ($r_0 \sim 0.5$ fm)
as  $a_{s}^{-1} =  1.100(6)$ GeV  ($a_s  \simeq 0.18$  fm). Hence  the
lattice  size  $12^3\times 96$  amounts  to $(2.15\mbox{fm})^3  \times
(4.30\mbox{fm})$ in the physical unit.

For quark  fields $\psi$ and $\bar\psi$, we  adopt the $O(a)$-improved
Wilson (clover) action on the anisotropic lattice as \cite{matsufuru}
\begin{eqnarray}
  S_{\rm F}
  &\equiv&
  \sum_{x,y} \bar\psi(x) K(x,y) \psi(y),
  \\\nonumber
  K(x,y)
  &\equiv&
  \renewcommand{\arraystretch}{1.8}
  \delta_{x,y}
  -
  \kappa_{t}\left\{\Tate\right.
  \Bs
  \begin{array}[t]{l} \displaystyle
    (1 - \gamma_4)\; U_4(x)\; \delta_{x+\hat 4,y}
    \\\displaystyle
    +
    (1 + \gamma_4)\; U_4^\dagger(x - \hat 4)\; \delta_{x-\hat 4,y}
    \left.\Tate\right\}
  \end{array}
  \\\nonumber
  &-&
  \renewcommand{\arraystretch}{1.8}
  \kappa_{s}
  \sum_i
  \left\{\Tate\right.
  \Bs
  \begin{array}[t]{l} \displaystyle
    (r - \gamma_i)\; U_i(x)\; \delta_{x+\hat i,y}
    \\\displaystyle
    +
    (r + \gamma_i)\; U_i^\dagger(x - \hat i)\; \delta_{x-\hat i,y}
    \left.\Tate \right\}
  \end{array}
  \\\nonumber
  &-&
  \kappa_{s}\; c_E \sum_i \sigma_{i4} F_{i4} \delta_{x,y}
  -
  r\; \kappa_{s}\; c_B \sum_{i>j} \sigma_{ij} F_{ij} \delta_{x,y},
\end{eqnarray}
where $\kappa_{s}$  and $\kappa_{t}$  denote the spatial  and temporal
hopping  parameters,  respectively.   $F_{\mu\nu}$ denotes  the  field
strength,  which  is  defined  through the  standard  clover-leaf-type
construction.
$r$  denotes the  Wilson  parameter. $c_{E}$  and  $c_{B}$ denote  the
clover coefficients.  
To achieve the tadpole improvement, the link variables are rescaled as
$U_i(x) \to  U_i(x)/u_s$ and $U_4(x) \to U_4(x)/u_t$,  where $u_s$ and
$u_t$ denote  the mean-field values  of the spatial and  temporal link
variables, respectively \cite{matsufuru,nemoto}.
This is  equivalent to the  redefinition of the hopping  parameters as
the   tadpole-improved   ones   (with   tilde),  i.e.,   $\kappa_s   =
\tilde\kappa_s / u_s$ and $\kappa_t = \tilde\kappa_t / u_t$.
The anisotropy parameter is defined as $\gamma_F \equiv \tilde\kappa_t
/ \tilde\kappa_s$,  which coincides with  the renormalized anisotropy
$\xi =  a_\sigma /  a_\tau$ for sufficiently  small quark mass  at the
tadpole-improved level \cite{matsufuru}.
For  given  $\kappa_s$, the  four  parameters  $r$,  $c_E$, $c_B$  and
$\kappa_s/\kappa_t$ should  be, in principle, tuned  so that ``Lorentz
symmetry'' holds up to discretization errors of $O(a^2)$.
Here, $r$, $c_E$ and $c_B$  are fixed by adopting the tadpole improved
tree-level values as
\begin{equation}
  r = \frac1{\xi},
  \Hs
  c_E = \frac1{u_\sigma u_\tau^2},
  \Hs
  c_B = \frac1{u_\sigma^3}.
\end{equation}
Only the  value of $\kappa_{\sigma} / \kappa_{\tau}  \left( = \gamma_F
\cdot (u_\sigma / u_\tau) \right)$ is tuned nonperturbatively by using
the meson dispersion relation \cite{matsufuru}.
It is convenient to define $\kappa$ as
\begin{equation}
  \frac1{\kappa}
  \equiv
  \frac1{\tilde\kappa_\sigma}
  -
  2\left( \gamma_F - 3r - 4 \right),
\end{equation}
then the bare  quark mass is expressed as  $m_0 = \frac1{2}(1/\kappa -
8)$ in the spatial lattice unit in the continuum limit.  This $\kappa$
plays the role of the  hopping parameter ``$\kappa$'' in the isotropic
formulation.
For detail, see Refs.~\cite{nemoto,matsufuru},  from which we take the
lattice  parameters.   The  values   of  the  lattice  parameters  are
summarized in \Table{table.lattice.parameters}.

We  adopt four  values  of the  hopping  parameter as  $\kappa=0.1210,
0.1220, 0.1230$ and $0.1240$,  which correspond to $m_{\pi}/m_{\rho} =
0.81, 0.77,  0.72$ and $0.65$, respectively.   For temporal direction,
we impose  anti-periodic boundary condition  on all the  quark fields.
For spatial directions, unless otherwise indicated, we impose periodic
boundary  condition on  all  the  quarks. We  refer  to this  boundary
condition as ``{\em standard (spatial) BC}''.

\begin{table}
\begin{ruledtabular}
\begin{tabular}{llllll}
$\kappa$   & 0.1210 & 0.1220 & 0.1230 & 0.1240 & $\kappa_{\rm phys.}$\\
\hline
$m_{\pi}$  & 1.005(2) & 0.898(2) & 0.784(2) & 0.656(3) & 0.140    \\
$m_{\rho}$ & 1.240(3) & 1.161(3) & 1.085(4) & 1.011(5) & 0.850(7) \\
$m_{K}$    & 0.845(2) & 0.785(2) & 0.723(2) & 0.656(3) & 0.530(4) \\
$m_{N}$    & 1.878(5) & 1.744(5) & 1.604(5) & 1.460(6) & 1.173(9)
\end{tabular}
\end{ruledtabular}
\caption{Masses of $\pi$,  $\rho$, K and N for  each hopping parameter
$\kappa$ in the physical  unit GeV.  $\kappa_{\rm phys.}\simeq 0.1261$
denotes  the value of  $\kappa$ which  achieves $m_{\pi}  \simeq 0.14$
GeV.}
\label{table.mass}
\end{table}
By  keeping  $\kappa_s=0.1240$ fixed  for  s  quark,  and by  changing
$\kappa=0.1210-0.1240$  for u  and  d quarks,  we  perform the  chiral
extrapolation to the physical quark mass region.
In the following part of the paper, we will use
\begin{equation}
  (\kappa_s,\kappa)=(0.1240,0.1220),
\label{typical.set}
\end{equation}
as a typical  set of hopping parameters in  presenting correlators and
effective mass plots.
For convenience,  we summarize  masses of $\pi$,  $\rho$, K and  N for
each   hopping  parameter  $\kappa$   together  with   their  chirally
extrapolated   values  in   \Table{table.mass}.    Here,  the   chiral
extrapolations of these particles are performed with a linear function
in $m_{\pi}^2$.
%
Unless  otherwise  indicated,   we  adopt  jackknife  prescription  to
estimate statistical errors.

In order  to enhance  the low-energy contributions,  we use  a smeared
source.    This  is   achieved  by   employing  a   spatially  extended
interpolating field  of the gaussian  size $\rho\simeq 0.4$ fm  in the
Coulomb gauge, which is obtained  by replacing the quark fields $q(x)$
in  \Eq{sasaki-op}  by the  smeared  quark  fields $q_{\rm  smear}(x)$
defined as
\begin{equation}
  q_{\rm smear}(t, \vec x)
  \equiv
  {\cal N}
  \sum_{\vec y}
  \exp\left\{ - \frac{|\vec x - \vec y|^2}{2\rho^2} \right\}
  q(t, \vec y),
\label{eq.gaussian}
\end{equation}
where ${\cal N}$ is  an appropriate normalization factor.
%
%
For practical  use, we extend \Eq{eq.gaussian} appropriately  so as to
fit a particular choice of the spatial boundary condition.
In  this paper, we  present correlators  with a  smeared source  and a
point sink.

\subsection{NK scattering states for standard BC}
The  NK scattering  state  with  zero total  momentum  is obtained  as
J=1/2-projection of the following state:
\begin{equation}
  \left|\Tate\right.
  N(\vec p,s) K(-\vec p) 
  \left.\Tate\right\rangle
  \pm
  \left|\Tate\right.
  N(-\vec p,s) K(\vec p) 
  \left.\Tate\right\rangle,
\label{eq.scattering.state}
\end{equation}
where ``$+$'' is  for s-wave, and ``$-$'' for  p-wave.
%
%
$s$ denotes the spin degrees of freedom of the nucleon.
In  finite spatial  box of  the size  $L^3$, the  allowed  momentum is
quantized as
\begin{equation}
  \vec p = \frac{2\pi}{L} \vec n,
  \Hs
  \vec n \in \ZZ^3,
\end{equation}
due to  the periodic boundary  condition.
%
%
By  neglecting the interaction  between N  and K,  the energy  of this
state is approximated as
\begin{equation}
  E(\vec p)
  \simeq
  \sqrt{m_K^2 + \vec p^2}
  +
  \sqrt{m_N^2 + \vec p^2}.
  \label{NK.scattering.energy}
\end{equation}
For s-wave, the NK threshold is  expressed as $E_{\rm th} \equiv m_K +
m_N$.  On  the other hand, for  p-wave, since \Eq{eq.scattering.state}
vanishes for $\vec  p = \vec 0$, the NK  threshold starts from $E_{\rm
th} \equiv  E(\vec p_{\rm min})$ associated with  the minimum momentum
\begin{equation}
  |\vec p_{\rm min}| = \frac{2\pi}{L}.
\label{eq.p-wave.minimum.p}
\end{equation}
Numerical  values of these  NK thresholds  for each  hopping parameter
$\kappa$ are presented in \Table{table.NK.thresholds}.
\begin{table}
\begin{ruledtabular}
\begin{tabular}{llllll}
$\kappa$ & 0.1210 & 0.1220 & 0.1230 & 0.1240 & empirical\\
\hline
s-wave & 2.723 & 2.528 & 2.327 & 2.116 & 1.440\\
p-wave & 2.987 & 2.809 & 2.629 & 2.442 & 1.865\\
HBC & 2.924 & 2.743 & 2.558 & 2.367 & 1.770\\
\end{tabular}
\end{ruledtabular}
\caption{Numerical values of NK thresholds (\Eq{NK.scattering.energy})
for each  hopping parameter  $\kappa$ in the  physical unit  GeV, when
spatial  lattice of  the size  is  $L\simeq 2.15$  fm.  The  rightmost
column  indicated  by  ``empirical''  corresponds  to  the  thresholds
calculated for  the physical masses of  N and K,  $m_{N}=0.94$ GeV and
$m_K=0.5$ GeV.
The first  and the second rows  show the NK thresholds  for the s-wave
and p-wave states, respectively, when the standard BC is imposed.  The
third row shows the NK thresholds for the HBC.  In the HBC, the s-wave
and p-wave NK thresholds coincide.}
\label{table.NK.thresholds}
\end{table}

\subsection{The correlators for $\Theta^+$}

\begin{figure}
\begin{center}
\includegraphics[height=0.48\textwidth,angle=-90]{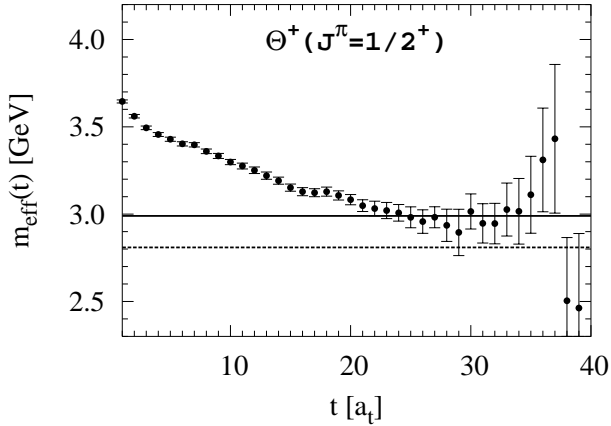}
\end{center}
\caption{The  effective  mass  plot  of  positive  parity  $\Theta^+$.
\Eq{typical.set} is  adopted as a  typical set of  hopping parameters.
The statistical error is  obtained with jackknife error estimate.
The  solid  line denotes  the  result  of  the single-exponential  fit
performed  over the  interval  $25 \le  t  \le 35$.   The dotted  line
denotes the  p-wave NK threshold  energy corresponding to  the spatial
lattice size $L\simeq 2.15$ fm.}
\label{fig.effmass.positive}
\end{figure}
In \Fig{fig.effmass.positive}, we show the effective mass plot for the
positive parity channel, which is  obtained from the correlator with a
smeared  source  and  a  point sink  adopting  \Eq{typical.set}.
%
%
The  dotted line  denotes  the  p-wave NK  threshold  for the  spatial
lattice size $L \simeq 2.15$ fm.

The effective mass is defined as
\begin{equation}
  m_{\rm eff}(t)
  \equiv
  \log\left({ G(t) \over G(t+1) }\right),
\label{eq.effmass}
\end{equation}
where $G(t)$ denotes the correlator.
At sufficiently large $t$, contributions from excited states diminish,
and the correlator  is dominated by a single state  with energy $m$ as
$G(t)  \sim  A  e^{-m  t}$.   Then \Eq{eq.effmass}  gives  a  constant
effective mass as $m_{\rm eff}(t) \sim m$.
Thus  a plateau  can be  served as  an indicator  of  the single-state
saturation. One is then  allowed to perform the single-exponential fit
in the plateau region.

In the  region $0  \le t \alt  25$ in  \Fig{fig.effmass.positive}, the
effective mass $m_{\rm eff}(t)$ decreases monotonically, which implies
that the higher spectral contributions are gradually reduced.
We  find a plateau  in the  interval $25  \alt t  \alt 35$,  where the
single-state dominance is expected to be achieved.
Beyond $t\sim  35$, the  data becomes too  noisy.  In addition,  it is
expected  to  receive  the  contribution  from  backwardly-propagating
negative parity state as mentioned before.
Hence, we  simply neglect the  data for $t  \agt 35$, and  perform the
single-exponential fit  of the correlator with  $f(t) = A  e^{-m t}$ in
the plateau  region, $25 \le  t \le 35$.   The solid line  denotes the
best-fit result.

\begin{figure}
\begin{center}
\includegraphics[height=0.48\textwidth,angle=-90]{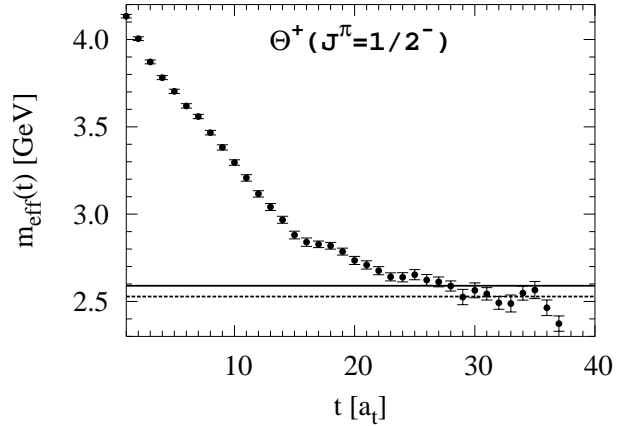}
\end{center}
\caption{The effective  mass plot  for negative parity  $\Theta^+$ for
the typical set of hopping parameters (\Eq{typical.set}).
The  solid  line denotes  the  result  of  the single-exponential  fit
performed  over the  interval  $25 \le  t  \le 35$.   The dotted  line
denotes the s-wave NK threshold.}
\label{fig.effmass.negative}
\end{figure}
In \Fig{fig.effmass.negative}, we show the effective mass plot for the
negative  parity channel  adopting \Eq{typical.set}.
The dotted line denotes s-wave NK threshold.
In the small $t$ region, i.e., $t \alt 25$, $m_{\rm eff}(t)$ decreases
monotonically,
%
and we see a rather stable plateau at $25 \alt t \alt 35$.
%
The single exponential fit at $25 \le t \le 35$ gives the solid line.

\begin{figure}
\begin{center}
\includegraphics[height=0.48\textwidth,angle=-90]{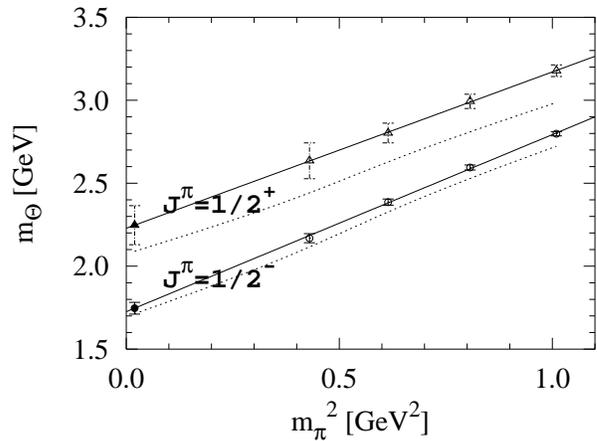}
\end{center}
\caption{$m_{\Theta}$   for  both   parity   states  plotted   against
$m_{\pi}^2$.  The triangles denote  positive parity, while the circles
denote negative  parity.  The open  symbols denote the  direct lattice
QCD data, whereas the closed  symbols denote the results of the chiral
extrapolation.   The  dotted  lines  indicate the  NK  thresholds  for
p-wave(upper) and s-wave(lower) cases.}
\label{fig.chiral.extrapolation}
\end{figure}
\begin{table}
\begin{ruledtabular}
\begin{tabular}{llllll}
$\kappa$ & 0.1210 & 0.1220 & 0.1230 & 0.1240 & $\kappa_{\rm phys.}$ \\
\hline
$+$ parity & 3.18(4) & 2.99(4) & 2.80(6) & 2.64(11) & 2.25(12)
\\
$-$ parity & 2.80(1) & 2.59(2) & 2.39(2) & 2.17(3) & 1.75(4)
\end{tabular}
\end{ruledtabular}
\caption{The  masses of  $\Theta^+$ for  each value  of  $\kappa$. The
first line  corresponds to the  positive parity state, and  the second
line to the negative  parity state.  $\kappa_{\rm phys.}\simeq 0.1261$
denotes the value of $\kappa$ which achieve $m_{\pi}\simeq 0.14$ GeV.}
\label{table.chiral.extrapolation}
\end{table}
As  mentioned  before, $\kappa=0.1240$  is  kept  fixed  for s  quark,
and $\kappa=0.1210-0.1240$  are varied for  u and d quarks  to perform
the  chiral  extrapolation.   In  \Fig{fig.chiral.extrapolation},  the
masses of positive (triangle)  and negative (circle) parity $\Theta^+$
are plotted  against $m_{\pi}^2$.  The open symbols  denote the direct
lattice data.  Since these data behave almost linearly in $m_{\pi}^2$,
the  chiral extrapolation  is performed  with a  linear  function.
Such  a  linear behavior  against  $m_{\pi}^2$  is  also observed  for
ordinary non-PS mesons and baryons \cite{nemoto}.
The closed  symbols denote the  results of chiral  extrapolation.  For
convenience, we  show p-wave (upper)  and s-wave (lower)  NK threshold
with dotted lines.

For  the positive  parity  state, the  chiral  extrapolation leads  to
$m_{\Theta}=2.25$ GeV,  which is much heavier  than the experimentally
observed $\Theta^+(1540)$.
For negative parity state, on the other hand, the chiral extrapolation
leads to $m_{\Theta}=1.75$ GeV, which is rather close to the empirical
value.  Our lattice QCD results  thus support that the negative parity
state is the lowest of the $\Theta^+$ spectrum.
\Table{table.chiral.extrapolation} summarizes  masses of $\Theta^+$ of
both parities for each  hopping parameter together with their chirally
extrapolated values.

\section{The new method to distinguish $\Theta^+$ and NK}
\label{section.hybrid.boundary.condition}
\subsection{The hybrid boundary condition (HBC)}
In  the p-wave NK  scattering states,  N and  K have  non-zero minimum
momenta as $|\vec{p}_{\rm min}| = 2\pi/L$ due to the finiteness of the
spatial box.  As  a consequence, the NK threshold is  raised by a few
hundred MeV depending on the spatial volume.
In studies of  positive parity $\Theta^+$, we can  utilize this volume
dependence  to  distinguish  localized resonances  from  NK-scattering
states.
In contrast,  in studies of negative parity  $\Theta^+$, NK scattering
states are of s-wave, and N  and K can have zero momenta, i.e., $|\vec
p_{\rm  min}|  =  0$.   Therefore,  the NK  threshold  has  no  volume
dependence, i.e.,  $m_{\rm N} +  m_{\rm K}$, which is  less convenient
to distinguish $\Theta^+$ from NK scattering states.

It would  be of  great worth,  if we could  find some  prescription to
raise the s-wave NK threshold by changing the spatial volume.
This can be  achieved by twisting the spatial  boundary condition in a
flavor dependent manner as follows.
We  impose the  anti-periodic boundary  condition on  u and  d quarks,
whereas   the  periodic   boundary   condition  on   s  quark.    (See
\Table{table.hybrid.bc}.)   We  will refer  to  this spatial  boundary
condition as ``{\em hybrid boundary condition (HBC)}''.
\begin{table}
\begin{ruledtabular}
\begin{tabular}{lccc}
  & u quark & d quark & s quark \\
\hline
HBC & anti-periodic & anti-periodic & periodic \\
standard BC & periodic & periodic & periodic
\end{tabular}
\end{ruledtabular}
\caption{The  {\em hybrid  boundary  condition (HBC)}  imposed on  the
quark fields.  The second line shows the standard BC for comparison.}
\label{table.hybrid.bc}
\end{table}

As a  consequence of the HBC,  hadrons feel the  boundary condition in
their own  ways.  Since N($uud$, $udd$)  and K($u\bar{s}$, $d\bar{s}$)
contain odd  numbers of the  u and d  quarks, they are subject  to the
anti-periodic     boundary    condition.     In     contrast,    since
$\Theta^+$($uudd\bar{s}$) contain  even numbers of u and  d quarks, it
is subject  to the periodic  boundary condition.  Recall that,  due to
the  finite size  of  the  lattice, the  allowed  spatial momenta  are
quantized as
\begin{equation}
  p_i =
  \left\{
  \renewcommand{\arraystretch}{1.5}
  \begin{array}{ll}
    \displaystyle
    2 n_i\pi/L & \mbox{for periodic BC} \\
    \displaystyle
    (2 n_i + 1)\pi/L & \mbox{for anti-periodic BC},
  \end{array}
  \right.
\end{equation}
with  $n_i \in  \ZZ$.   Hence, N  and  K acquire  the non-zero  minimum
momenta as
\begin{equation}
  |\vec p_{\rm  min}| = \frac{\sqrt{3}\pi}{L},
  \label{eq.s-wave.minimum.p}
\end{equation}
even    for   the    s-wave   state.     By   comparing    this   with
\Eq{eq.p-wave.minimum.p}, we  see that HBC can raise  the NK threshold
like   the   p-wave   case.    (See   \Eq{eq.p-wave.minimum.p}.)    In
\Table{table.NK.thresholds}, we  also show NK thresholds  for the HBC.
Note that,  unlike the  standard BC, s-wave  threshold and  p-wave one
coincide in the HBC.
We see,  in \Table{table.NK.thresholds},  that s-wave NK  threshold is
raised by about  $200-250$ MeV in the range of  $0.1210 \le \kappa \le
0.1240$ in our calculation.

On the  other hand, a localized  resonance $\Theta^+(uudd\bar{s})$ can
have   zero   momentum   as   $|\vec   p_{\rm  min}|   =   0$.    (See
\Table{table.hadron.hybrid.bc}.)
Therefore, the shift of $m_{\Theta}$  comes only through the change in
its intrinsic  structure.  In this case,  the shift is  expected to be
less significant  than the shift due  to the kinematic  reason as is
the case in N and K.
%
Now, if  the spatial size  of $\Theta^+$ is sufficiently  smaller than
the spatial lattice size $L$, then it is safely expected that the mass
of $\Theta^+$ is insensitive to the change of the boundary condition.
\begin{table*}
\begin{ruledtabular}
\begin{tabular}{lccll}
	& quark content & spatial BC & minimum momentum &\\
\hline
N &$uud$, $udd$ & anti-periodic
		& $\vec p_{\rm min} = (\pm\frac{\pi}{L},\pm\frac{\pi}{L},\pm\frac{\pi}{L})$ &
		$|\vec p_{\rm min}| = \sqrt{3}\frac{\pi}{L}$ \\
K & $u\bar{s}$, $d\bar{s}$ & anti-periodic
		& $\vec p_{\rm min} = (\pm\frac{\pi}{L},\pm\frac{\pi}{L},\pm\frac{\pi}{L})$ &
		$|\vec p_{\rm min}| = \sqrt{3}\frac{\pi}{L}$ \\
$\Theta^+$ & $uudd\bar{s}$ & periodic
		& $\vec p_{\rm min} = (0,0,0)$ &
		$|\vec p_{\rm min}| = 0$ \\
\end{tabular}
\end{ruledtabular}
\caption{The consequence of the HBC on the hadrons.}
\label{table.hadron.hybrid.bc}
\end{table*}

\subsection{Numerical result on substance of 5Q state}
\label{section.numerical.result.ii}

\begin{figure}
\begin{center}
\includegraphics[height=0.45\textwidth,angle=-90]{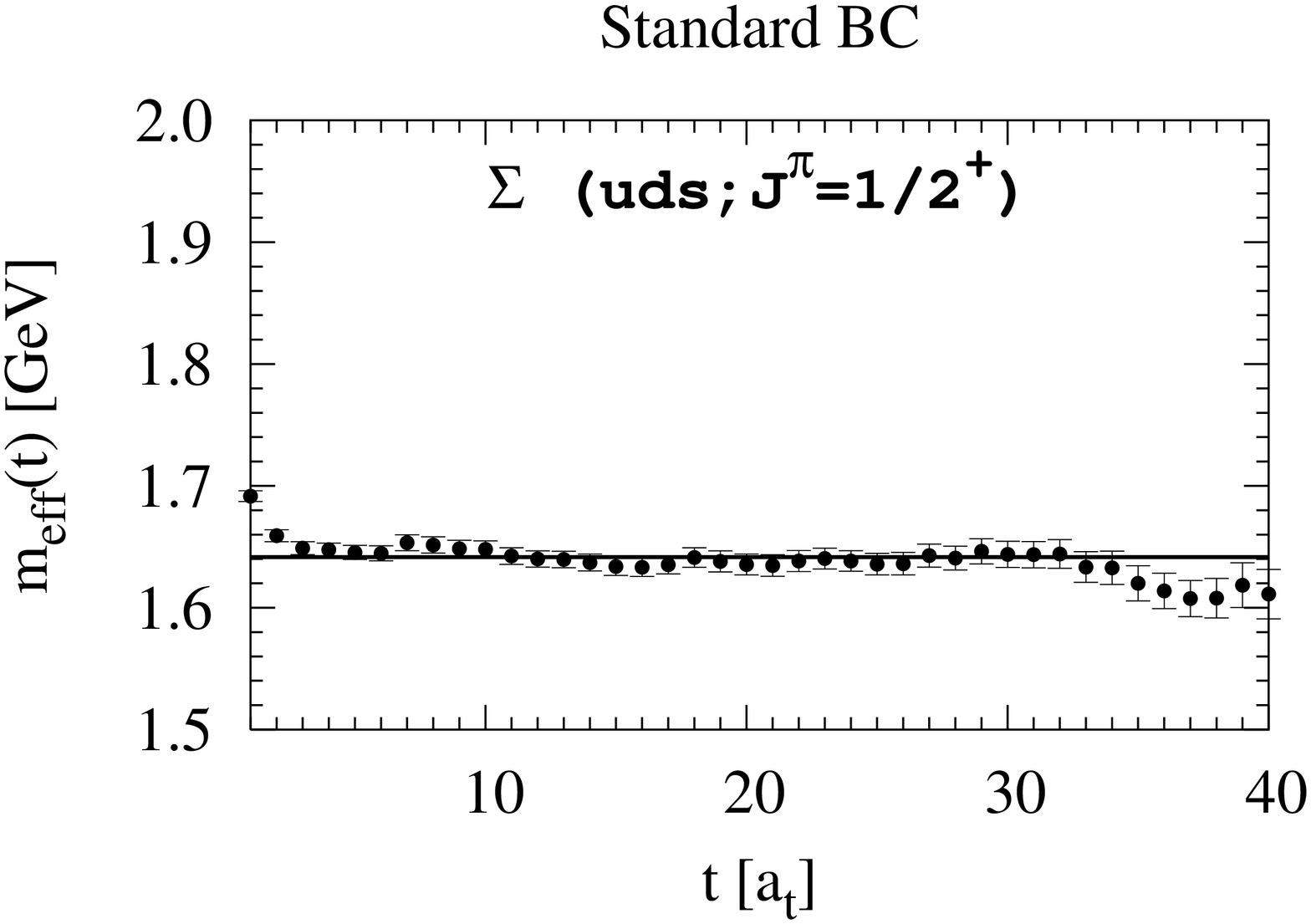}
\\
\includegraphics[height=0.45\textwidth,angle=-90]{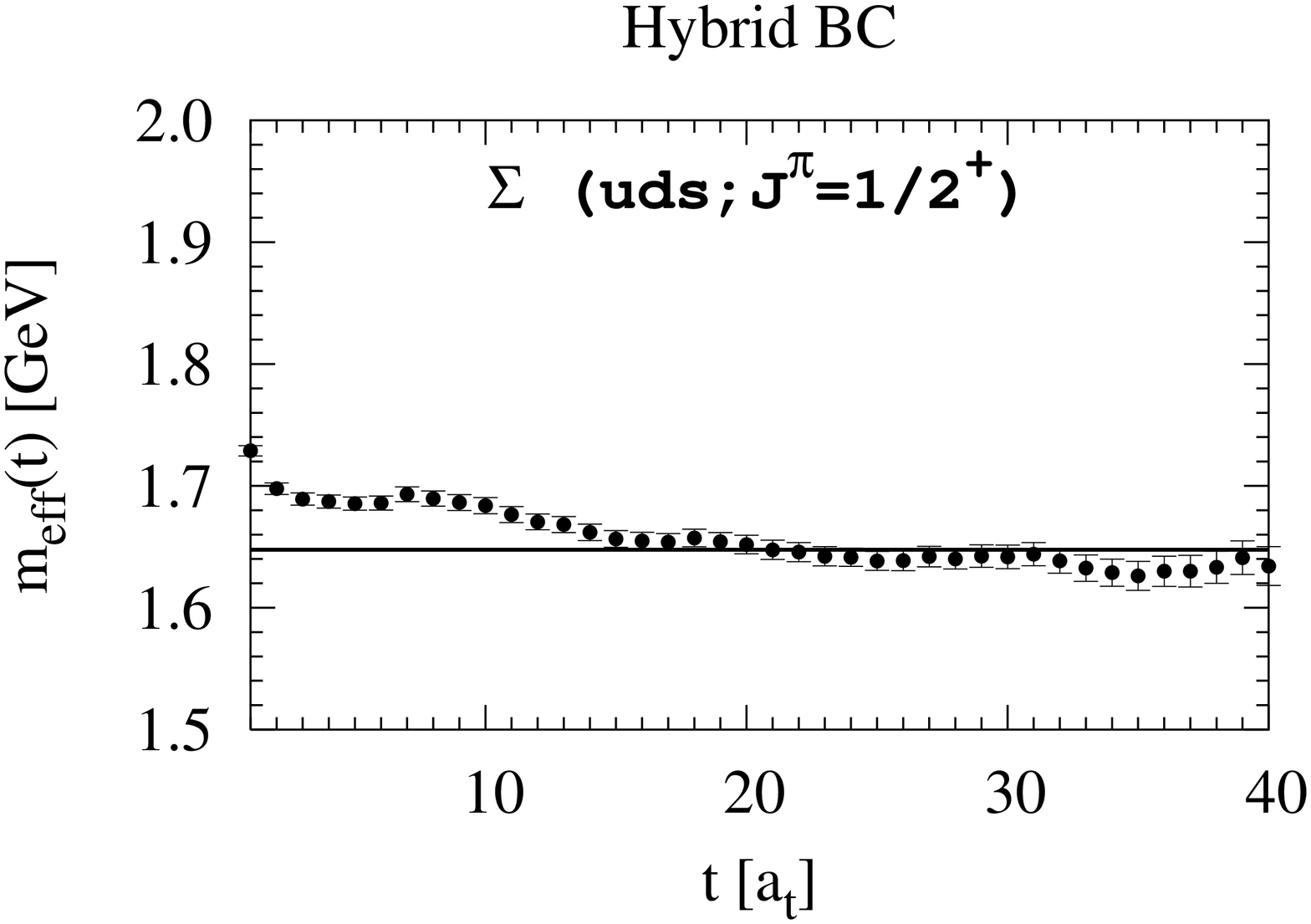}
\end{center}
\caption{The effective mass plots  of $\Sigma$($uds$) baryon under the
standard  BC, and  under the  HBC.  \Eq{typical.set}  is adopted  as a
typical set  of the  hopping parameters.  The  solid lines  denote the
results of the single exponential  fit performed in the region $20 \le
t \le 30$.
No significant difference is observed between the two results.}
\label{fig.sigma}
\end{figure}
Before  presenting the  results for  $\Theta^+$, we  have  to convince
ourselves that  localized resonance  states are really  insensitive to
the change  of boundary  condition.
For  this purpose,  we apply  the HBC  to the  $\Sigma$($uds$) baryon,
which is an established localized resonance.
%
We take  $\Sigma$ rather than  the nucleon, because  $\Sigma$ contains
even numbers of $u$ and $d$ quarks.  Hence, $\Sigma$ is subject to the
periodic boundary condition, and can have zero spatial momentum.
Therefore, if  the mass of $\Sigma$  is affected by the  choice of the
BC, it is attributed to a change of intrinsic structure.
%
%
%
In \Fig{fig.sigma},  we show effective  mass plots of  $\Sigma$ baryon
correlators   adopting   the   typical   set  of   hopping   parameter
\Eq{typical.set}.  The  upper one corresponds to the  standard BC, and
the lower one to the HBC. We  see that the plateau is raised by only a
negligible amount due  to the change of the  boundary condition.  This
example   explicitly  shows  that   localized  resonance   states  are
insensitive to the change of the boundary condition.

\begin{figure}
\begin{center}
\includegraphics[height=0.48\textwidth,angle=-90]{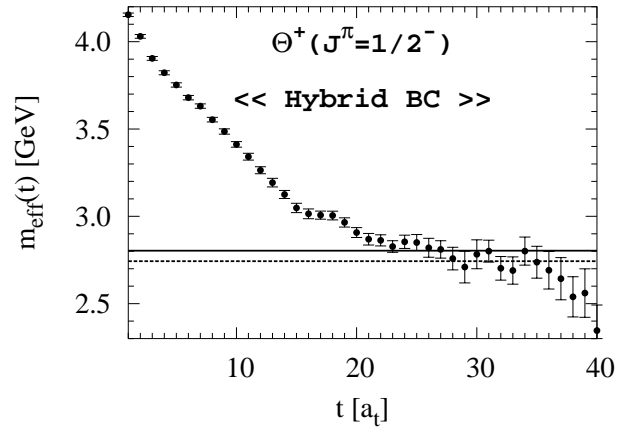}
\end{center}
\caption{The effective  mass plot  for the negative  parity $\Theta^+$
under  the HBC.   \Eq{typical.set}  is  adopted as  a  typical set  of
hopping parameters.
The  solid  line denotes  the  result  of  the single-exponential  fit
performed  within the interval  $25 \le  t \le  35$.  The  dotted line
denotes the s-wave NK threshold  energy modified due to the HBC.  This
figure should be compared with \Fig{fig.effmass.negative}.}
\label{fig.hybrid}
\end{figure}

Now, we  present the numerical  result of negative  parity $\Theta^+$.
In  \Fig{fig.hybrid}, we show  the effective  mass of  negative parity
$\Theta^+$   imposing   the   HBC   adopting   $\kappa_s=0.1240$   and
$\kappa=0.1220$.
In  the   region  $0<  t   \alt  25$,  the  effective   mass  decrease
monotonically as before.  There is a  plateau in the region $25 \alt t
\alt 35$.   Beyond $t  \sim 35$,  the plateau breaks  down due  to the
contributions    from     backwardly-propagating    positive    parity
states. Hence, we simply neglect  the data for $t\alt 35$, and perform
the  single-exponential fit in  the plateau  region, $25\le  t\le 35$.
The solid line denotes the  result of the single-exponential fit.  The
dotted  line denotes  the modified  NK threshold,  i.e., $\sqrt{m_{\rm
N}^2 + \vec p_{\rm min}^2}  + \sqrt{m_{\rm K}^2 + \vec p_{\rm min}^2}$
with $|\vec p_{\rm min}| = \sqrt{3}\pi/L$ due to the HBC.
\begin{figure}
\begin{center}
\includegraphics[height=0.48\textwidth,angle=-90]{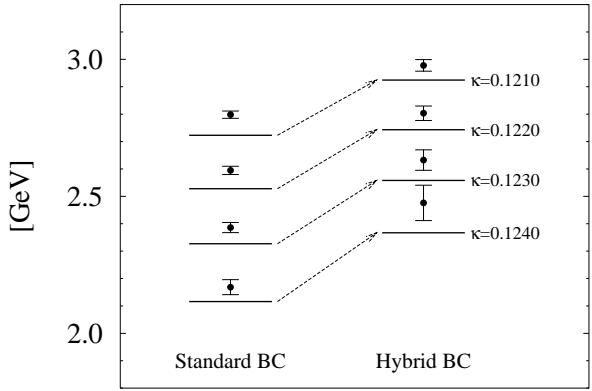}
\end{center}
\caption{Comparison of results of standard BC(l.h.s.)  and HBC(r.h.s.)
for each  hopping parameter  $\kappa$.  Closed circles  denote results
obtained  from the  best  fit  analysis. The  solid  lines denote  the
corresponding NK threshold.}
\label{fig.hikaku}
\end{figure}
Comparing \Fig{fig.hybrid} with \Fig{fig.effmass.negative}, we observe
that the plateau is raised by  about 200 MeV, which coincides with the
shift of the NK threshold.
In  \Fig{fig.hikaku},  we compare  the  result  of  HBC with  that  of
standard  BC for  each  hopping parameter  $\kappa$  for the  negative
parity $\Theta^+$.   Closed circles  denote the results  obtained with
best-fit   analysis.    Solid   lines   show  the   corresponding   NK
thresholds. For all $\kappa$, we see that the best-fit masses($\simeq$
the location of  the plateaus) are raised by about  the same amount as
the corresponding NK thresholds.
Therefore, we find  that there is no localized  resonance states below
$\sqrt{m_{\rm N}^2  + \vec p_{\rm  min}^2} + \sqrt{m_{\rm K}^2  + \vec
p_{\rm  min}^2}$  with $|\vec  p_{\rm  min}|  =  \sqrt{3}\pi/L$. As  a
result, we conclude that the  negative parity state observed before is
an NK scattering state.

\subsection{Comments}
(i) The fully anti-periodic  boundary condition is less convenient for
our purpose.
In  fact,  in  this  case,  K  is subject  to  the  periodic  boundary
condition, whereas  N and $\Theta^+$ are subject  to the anti-periodic
boundary condition.  Since K can  have the zero momentum, the shift of
the  NK  threshold is  smaller.   In  addition,  $\Theta^+$ must  have
non-zero momentum, which raises the  location of $\Theta^+$ due to the
kinematic  reason.  It  is  easy to  understand  that the  resulting
spectrum is less convenient for our purpose.

(ii) It is possible to use the HBC in the study of the positive parity
state.  However,  since the change  of the minimum momenta  amounts to
only  a  minor  modification  from $2\pi/L$  to  $\sqrt{3}\pi/L$,  the
resulting change in the spectrum is less significant than the negative
parity state.

(iii) The HBC provides us with a general tool, the use of which is not
restricted to the calculation of  $\Theta^+$.  For instance, it can be
used in the  study of the hadronic molecules.   Hadronic molecules are
bound state of two or more hadrons. The binding energies are typically
$10-50$ MeV.  In  the lattice QCD studies, the gap  of $10-50$ MeV may
be too small  to confirm the existence of  the bound state.  Therefore
the HBC is useful by enlarging this gap to a few hundred MeV.
\\

(iv) Careful  reader may  point out that,  besides the  stable plateau
which  we  discussed,  there  is  a small  plateau-like  structure  in
\Fig{fig.effmass.negative} and \Fig{fig.hybrid}  in the region $15 \le
t \le 18$.
However, this ``plateau''  is also raised by about 200  MeV due to the
HBC. This means that this region is not dominated by a single state,
and a  possible reason of this  shift would be  that contaminations of
low-lying states becomes reduced in the HBC.
Still, the nature of this  small plateau cannot be determined from our
data available so far.
At  any  rate,  we can  state  that  it  does  not correspond  to  the
experimentally  observed  $\Theta^+(1540)$,  because its  location  is
rather high $m\simeq 1.95$ GeV even after the chiral extrapolation.
Note  that  we  do not  find  such  a  plateau-like structure  in  the
effective mass plot with a point source and a point sink.

\section{Summary and discussion}
\label{section.summary.and.discussion}

We  have  studied the  penta-quark  baryon  $\Theta^+$ in  anisotropic
lattice QCD at the quenched level. We have used the standard plaquette
action for  gauge field configurations  on the anisotropic  lattice of
the size  $12^3\times 96$ with the  renormalized anisotropy $a_s/a_t=4$
at  $\beta=5.75$, i.e., $a_s  \simeq 0.18$  fm, $a_t\simeq  0.045$ fm.
%
For  quarks, we  have adopted  $O(a)$-improved Wilson  (clover) action
with   four    values   of    hopping   parameters   as    $\kappa   =
0.1210(0.0010)0.1240$.
To enhance the low-lying spectra, we have adopted a spatially extended
operator.

We  have  used  504   gauge  field  configurations  to  construct  the
correlator  of a  non-NK type  interpolating field.   For  each parity
channel,  we have  found a  rather  stable plateau,  where the  single
exponential fit has been performed. After the chiral extrapolation, we
have found  $m_{\Theta}=2.25$ GeV for positive parity  state, which is
too   heavy   to   be   accepted  as   the   experimentally   observed
$\Theta^+(1540)$.   For   negative  parity  case,   we  have  obtained
$m_{\Theta}=1.75$  GeV,  which  is   rather  close  to  the  empirical
value. Our data have thus  suggested that the negative parity state is
the lowest of the penta-quark spectrum.

In order to clarify whether  this negative parity state is a localized
resonance or  not, we  have proposed a  new method with  ``{\em hybrid
boundary condition (HBC)}'',  where the anti-periodic spatial boundary
condition is imposed on the u and d quarks, while the periodic spatial
boundary condition  is imposed  on the s  quark.  As  results, $N(uud,
udd)$ and  $K(u\bar{s}, d\bar{s})$, both of which  contain odd numbers
of u and  d quarks, are subject to  the anti-periodic spatial boundary
condition,   whereas  $\Theta^+(uudd\bar{s})$,  which   contains  even
numbers of  the u and  d quarks, is  subject to the  periodic boundary
condition.   A remarkable feature  of the  HBC is  that it  raises the
s-wave  NK  threshold by  a  few  hundred  MeV without  affecting  the
localized penta-quark resonance.

By using  the HBC  method, we have  further investigated  the negative
parity state.
We have found that  the plateau is raised by about 200  MeV due to the
HBC,  which is  consistent with  the shift  of the  NK  threshold.  We
conclude   that  there   is   no  localized   resonance  state   below
$\sqrt{m_{\rm N}^2  + \vec p_{\rm  min}^2} + \sqrt{m_{\rm K}^2  + \vec
p_{\rm  min}^2}$  with  $|\vec  p_{\rm  min}|  =  \sqrt{3}\pi/L$.   It
follows, in particular, that the negative parity state observed in the
standard BC is a mere NK scattering state.

Now we discuss possible implications of our results.
Experimentally,          a          number          of          groups
\cite{nakano,diana,clas,saphir,experiments}    have    confirmed   the
existence of $\Theta^+(1540)$. Hence, it should be observed in lattice
QCD Monte Carlo calculation as well.
However, in neither positive nor negative parity states, we have found
any relevant signals of narrow penta-quark states.
%
In   this   way,   we   have   arrived  at   a   similar   result   as
\Ref{kentacky}. However,  it should be  remarked that this  result has
been obtained in spite of  our choice of the non-NK type interpolating
field.
%
These two  results suggest that $\Theta^+$  may not be  reachable by a
simple five-quark  interpolating field  operator in lattice  QCD ---it
might have more complicated structure than expected.
%
Needless to say, it is desirable to examine the finite volume artifact
and discretization  error, as usual.   It is also desirable  to reduce
the ambiguity from the chiral extrapolation.
Concerning  the  light  quark  effects,  one  might  wonder  that  the
dynamical  quark  effects  might  be  important,  because  the  narrow
resonance  $\Theta^+$ was  originally  predicted by  the Skyrme  model
\cite{diakonov}.
After  all,  the  Skyrmion  is  a nontrivial  configuration  of  pion.
However, this argument is not  quite correct, since the nucleon serves
as a counter-example.
A nucleon is  indeed the ground state of in the  Skyrme model and thus
contains  a non-trivial  configuration of  pion.  It  is  however also
fairly  well  reproduced in  quenched  lattice  QCD with  conventional
three-quark interpolating field.

%
Besides these rather technical stuff, the difficulty may trace back to
our insufficient understanding of  the true nature of the penta-quark.
It would be interesting to consider the following possibilities.

(1) Highly  non-local structure:  If  the structure  of $\Theta^+$  is
    really the Jaffe-Wilczek type  or the Karliner-Lipkin type, it may
    be better to invent  series of new non-local interpolating fields,
    which have  sufficiently complicated  structures so that  they can
    fit Jaffe-Wilczek/Kaliner-Lipkin  pictures.  In this  respect, the
    group theoretical construction of interpolating fields proposed in
    \Ref{morningstar}  together with the  variational method  would be
    interesting.

(2) Other  quantum  numbers:  In  most  lattice  QCD  calculations  of
    $\Theta^+$, spin of $\Theta^+$ is assumed to be 1/2.  Naively, one
    may expect  that J=3/2 state  would be heavier than  J=1/2 states.
    However, since  there is no  experimental evidence on the  spin of
    $\Theta^+$, it would be  worth keeping J=3/2 and 5/2 possibilities
    as candidates.  We have to keep in mind that $\Theta^+$ is already
    a quite mysterious particle.
    In addition, although  there is no evidence on  the doubly charged
    $\Theta^{++}$ in  the experiment \cite{saphir}, it  might be worth
    studying I=1  and 2 states as  well, which have  been suggested by
    \Ref{capstick}. Since  we do not know the  production mechanism of
    the  penta-quark, it  might be  better  to keep  them as  possible
    candidates.

(3) $K\pi N$ hepta-quark hypothesis:  All the lattice QCD calculations
    for $\Theta^+$ available so far  are based on the quenched QCD. It
    is known that  some hadrons are difficult to  be reproduced in the
    quenched QCD. For instance,  quenched lattice QCD cannot reproduce
    $\Lambda(1405)$  as  a  three-quark  state \cite{nemoto}.   It  is
    conjectured  to  be  a  $\bar{K}N$  bound state.   In  this  case,
    $\Lambda(1405)$ is  a ``penta-quark''. Hence, it  would be natural
    that  mass of  $\Lambda(1405)$  cannot be  properly reproduced  in
    quenched  lattice QCD  with a  standard  three-quark interpolating
    field.  Actually, $\Theta^+(1540)$ itself  is hypothesized to be a
    bound    state    of    $K\pi    N$,   i.e.,    the    hepta-quark
    \cite{bicudo,kishimoto,oset}.
    If  this   is  really   the  case,  it   would  be   natural  that
    $\Theta^+(1540)$ is  difficult to be observed  in quenched lattice
    QCD with  ``ordinary'' penta-quark interpolating fields.
    However,  concerning the  hepta-quark picture,  we should  keep in
    mind that,  in spite  of its many  good features,  the interaction
    among $K\pi N$ system is said to be too weak to make a bound state
    with the binding energy of 40 MeV.

In this way, it is necessary to perform more systematic studies of the
penta-quarks in order to reveal its mysterious nature.
The proper use of lattice QCD can help us proceed in this direction.

\begin{acknowledgments}
We  thank  H.~Matsufuru, Y.~Nemoto,  T.T.~Takahashi  and T.~Umeda  for
useful information and discussions.
%
%
M.~O and H.~S  are supported in part by  Grant for Scientific Research
((B)  No.   15340072 and  (C)  No.   16540236)  from the  Ministry  of
Education, Culture, Sports, Science and Technology, Japan.
T.~D. is supported by Special Postdoctoral Research Program of RIKEN.
H.~I. is supported by a 21st  Century COE Program at Tokyo Institute of
Technology  ``Nanometer-Scale  Quantum Physics''  by  the Ministry  of
Education, Culture, Sports, Science and Technology.
%
%
The  lattice  QCD Monte  Carlo  calculations  have  been performed  on
NEC-SX5 at Osaka University.
\end{acknowledgments}

\appendix
\section{rearrangements of Eq.~(\ref{sasaki-op})}
\label{appendix}
The aim here is to  estimate the contribution from NK-type operator to
the  correlator  of  \Eq{sasaki-op}.
To this end, we use the Fierz rearrangements.
By     using    an     identity     $\epsilon_{abc}\epsilon_{bfg}    =
\delta_{cf}\delta_{ag} - \delta_{cg}\delta_{af}$,  and by applying the
Fierz  identity  to  $\left(  u_f^T  C  d^g  \right)  C  \bar{s}_c^T$,
\Eq{sasaki-op} is rearranged as
\begin{eqnarray}
  \lefteqn{
    \epsilon_{abc}\epsilon_{ade}\epsilon_{bfg}
    \left( u^T_d C\gamma_5 d_e \right)
    \left( u^T_f C d_g \right)
    C \bar{s}^T_c
  }
  \label{eq.fierz.1}
  \\\nonumber
  &=&
  \epsilon_{ade}
  \left( u^T_d C\gamma_5 d_e \right)
  \left\{
  \left( u^T_c C d_a \right) C \bar{s}^T_c
  -
  \left( u^T_a C d_c \right) C \bar{s}^T_c
  \right\}
  \\\nonumber
  &=&
  \epsilon_{ade}
  \left( u^T_d C\gamma_5 d_e \right)
  \\\nonumber
  &&
  \times
  \frac1{4}
  \left\{\Tate\right.
  - \gamma_5 d_a \left( \bar{s}_c\gamma_5 u_c \right)
  + \gamma_5 u_a \left( \bar{s}_c\gamma_5 d_c \right)
  \\\nonumber
  &&
  - d_a \left( \bar{s}_c u_c \right)
  + u_c \left( \bar{s}_c d_c \right)
  \\\nonumber
  &&
  + \gamma_{\mu} d_a \left( \bar{s}_c\gamma_{\mu} u_c \right)
  - \gamma_{\mu} u_a \left( \bar{s}_c\gamma_{\mu} d_c \right)
  \\\nonumber
  &&
  + \gamma_5\gamma_{\mu} d_a \left( \bar{s}_c\gamma_5\gamma_{\mu} u_c \right)
  - \gamma_5\gamma_{\mu} u_a \left( \bar{s}_c\gamma_5\gamma_{\mu} d_c \right)
  \\\nonumber
  &&
  + \frac1{2} \sigma_{\mu\nu} d_a \left( \bar{s}_c\sigma_{\mu\nu} u_c \right)
  - \frac1{2} \sigma_{\mu\nu} u_a \left( \bar{s}_c\sigma_{\mu\nu} d_c \right)
  \left.\Tate\right\},
\end{eqnarray}
where only the first and the second term factorize into the product of
the interpolating fields of N and K up to a factor of $\gamma_5$.
%
We see  that \Eq{sasaki-op}  contains the NK-type  interpolating field
with a factor of $\frac1{4}$.
%

In \Eq{eq.fierz.1},  u and d fields  in K come  from the pseudo-scalar
diquark operator.
There is another contribution, where u and d fields in K come from the
scalar  diquark  operator. This  is  obtained  by  using the  identity
$\epsilon_{abc}\epsilon_{ade}      =      \delta_{bd}\delta_{ce}     -
\delta_{be}\delta_{cd}$  and   by  applying  the   Fierz  identity  to
$\left(u^T_d C\gamma_5 d_3\right) C \bar{s}_c^T$ in the following way:
\begin{eqnarray}
  \lefteqn{
    \epsilon_{abc}\epsilon_{ade}\epsilon_{bfg}
    \left( u^T_d C\gamma_5 d_e \right)
    \left( u^T_f C d_g \right)
    C \bar{s}^T_c
  }
  \label{eq.fierz.2}
  \\\nonumber
  &=&
  \epsilon_{bfg}
  \left( u^T_f C d_g \right)
  \left\{
  \left( u^T_b C\gamma_5 d_c \right) C \bar{s}^T_c
  -
  \left( u^T_c C\gamma_5 d_b \right) C \bar{s}^T_c
  \right\}
  \\\nonumber
  &=&
  \epsilon_{bfg}
  \left( u^T_f C d_g \right)
  \\\nonumber
  &&
  \times
  \frac1{4}
  \left\{\Tate
  - u_b \left( \bar{s}_c \gamma_5 d_c \right)
  + d_b \left( \bar{s}_c \gamma_5 u_c \right)
  \right.
  \\\nonumber
  &&
  - \gamma_5 u_b \left( \bar{s}_c d_c \right)
  + \gamma_5 d_b \left( \bar{s}_c u_c \right)
  \\\nonumber
  &&
  - \gamma_5\gamma_{\mu} u_b \left( \bar{s}_c \gamma_{\mu} d_c \right)
  + \gamma_5\gamma_{\mu} d_b \left( \bar{s}_c \gamma_{\mu} u_c \right)
  \\\nonumber
  &&
  - \gamma_{\mu} u_b \left( \bar{s}_c \gamma_5\gamma_{\mu} d_c \right)
  + \gamma_{\mu} d_b \left( \bar{s}_c \gamma_5\gamma_{\mu} u_c \right)
  \\\nonumber
  &&
  \left.
  + \frac1{2} \sigma_{\mu\nu} u_b \left( \bar{s}_c \sigma_{\mu\nu} d_c \right)
  - \frac1{2} \sigma_{\mu\nu} d_b \left( \bar{s}_c \sigma_{\mu\nu} u_c \right)
  \right\},
\end{eqnarray}
where only the first and the second terms contribute to the product of
interpolating fields  of N  and K up to a factor of $\gamma_5$.
The nucleon interpolating field  here has an unfamiliar form.  Another
Fierz rearrangement of this  nucleon interpolating field can make this
into  the  familiar  form,  which  provides an  additional  factor  of
$\frac1{4}$.  Totally,  the contribution of  the NK-type interpolating
field to \Eq{sasaki-op} in this case involves the factor of 1/16.
%

By   neglecting  the   small  contribution   in   \Eq{eq.fierz.2}  for
simplicity,
the direct contribution of  the NK-type interpolating field (in source
and sink) to  the correlator of \Eq{sasaki-op} involves  the factor of
1/16, which is rather small.

\end{document}